\documentclass[%
 superscriptaddress,
reprint,
 amsmath,amssymb,
 aps,longbibliography
]{revtex4-1}

\usepackage{graphicx}
\usepackage{dcolumn}
\usepackage{bm}
\usepackage{hyperref}
\usepackage[mathlines]{lineno}
\usepackage[bottom]{footmisc}
\usepackage{booktabs}
\makeatletter
\renewcommand*{\@fnsymbol}[1]{\ensuremath{\ifcase#1\or *\or ** \or \ddagger\or
    \mathsection\or \mathparagraph\or \|\or **\or \dagger\dagger
    \or \ddagger\ddagger \else\@ctrerr\fi}}
\makeatother

\begin{document}

\title{Phase-modulated axilenses as ultra-compact spectroscopic tools }

\author{Wesley A. Britton}
\thanks{These authors contibute equally to this work}
\affiliation{Division of Materials Science \& Engineering, Boston University, 15 Saint Mary's St., Brookline, Massachusetts 02446, USA}

\author{Yuyao Chen}
\thanks{These authors contibute equally to this work}
\affiliation{Department of Electrical \& Computer Engineering and Photonics Center, Boston University, 8 Saint Mary's Street, Boston, Massachusetts, 02215, USA}

\author{Fabrizio Sgrignuoli}
\affiliation{Department of Electrical \& Computer Engineering and Photonics Center, Boston University, 8 Saint Mary's Street, Boston, Massachusetts, 02215, USA}

\author{Luca Dal Negro}
\email[email:]{dalnegro@bu.edu}
\affiliation{Division of Materials Science \& Engineering, Boston University, 15 Saint Mary's St., Brookline, Massachusetts 02446, USA}
\affiliation{Department of Electrical \& Computer Engineering and Photonics Center, Boston University, 8 Saint Mary's Street, Boston, Massachusetts, 02215, USA}
\affiliation{Department of Physics, Boston University, 590 Commonwealth Avenue, Boston, Massachusetts, 02215, USA}

\begin{abstract}
We design and characterize phase-modulated, ultra-compact, silicon-based axilens devices that combine efficient point focusing and grating selectivity within scalable 4-level phase mask configurations. The proposed designs are polarization insensitive and maintain a large focusing efficiency over a broad spectral band. Specifically, we select and systematically characterize structures designed for visible and near-infrared (NIR) operation in the $750nm-950nm$ wavelength range. These devices are ideally suited for monolithic integration atop the substrate layers of focal plane arrays (FPAs) for use in multi-band photo-detection and imaging. We demonstrate linear control of multi-wavelength focusing on a single achromatic plane and provide an application consisting of a proof-of-concept ultra-compact single-lens spectrometer with only 300 nm thickness and 70${\mu}m$ diameter, achieving a minimum distinguishable wavelength  $\Delta\lambda=40nm$ at $\lambda_0=850nm$. The proposed devices add fundamental spectroscopic capabilities to compact imaging devices for a number of applications ranging from spectral sorting to visible and NIR multispectral imaging and detection. 
\end{abstract}

\maketitle
Focal plane arrays (FPAs) are central to modern imaging and sensing technology \cite{krishna2007quantum,stiff2009quantum,choi2017resonant,zhang2018solid}. Recently, microlenses have been proposed to work as optical concentrators in combination with FPAs in order to increase their sensitivity and suppress crosstalks  \cite{chen2002monolithic,jian2012design,bai2014performance,akin2015mid,allen2016increasing}. Specifically, microlenses based on metasurfaces (i.e. metalenses) have received significant attention due to their compact size and optically flat profiles  \cite{arbabi2015subwavelength,khorasaninejad2016polarization,chen2019broadband}. However, traditional metalens designs achieve desired phase profiles either through engineered resonance behavior, which introduces unavoidable optical losses reducing their overall focusing efficiency \cite{zhan2016low,khorasaninejad2016polarization}, or through geometrical phase modulation, which requires polarization control \cite{lin2014dielectric,zheng2015metasurface}.  In addition, controlled fabrication of advanced metasurfaces often requires sub-wavelength accuracy adding to cost and device complexity \cite{arbabi2016multiwavelength,chen2018broadband,shrestha2018broadband,yuan2017achromatic}.

In this letter, we propose an alternative microlens design approach based on the spatial modulation of ultra-compact, phase-modulated axilenses \cite{davidson1991holographic} in silicon that can be readily integrated atop the substrates of IR-FPAs. These devices efficiently integrate the characteristic phase modulation of axilenses, i.e. diffractive lenses with radially-dependent focal lengths and controllable depth of focusing, with the angular dispersion of grating structures. Building on these principles, we demonstrate novel 4-level diffractive optical elements (DOEs) with combined imaging and spectroscopic functionalities that are fully compatible with top-down fabrication processing and scalable photolithography. Notably, the proposed approach is polarization insensitive and can be extended to any desired spectral range. Consequently, the resulting devices enable  simultaneous control of the spatial and spectral information encoded in the incident radiation that are important to a number of on-chip spectroscopic and imaging applications.

In what follows, we study the wave diffraction problem of 4-level phase-modulated axilenses using the rigorous Rayleigh-Sommerfeld (RS) first integral formulation \cite{goodman2005introduction}. While the proposed approach is general, we will focus for simplicity on two-dimensional (2D) periodic phase modulations within a 4-level mask geometry with a maximum diameter $D=70{\mu}m$ and $\approx$ 300 nm thickness, which is a size comparable with typical metalens approaches \cite{arbabi2015subwavelength,chen2018broadband}. Secondly, we show that engineered phase-modulated axilenses can tightly focus incoming radiation with efficiency $\sim{35}\%$  on the same achromatic plane at different transverse locations linearly related to the incoming radiation wavelength. The angular dispersion characteristics of the fabricated devices are experimentally measured and a proof-of-concept, ultra-compact single-lens spectrometer is demonstrated and utilized to measure the emission spectrum of a light-emitting diode (LED).

\begin{figure*}[t]
\centering
\includegraphics[width=\linewidth]{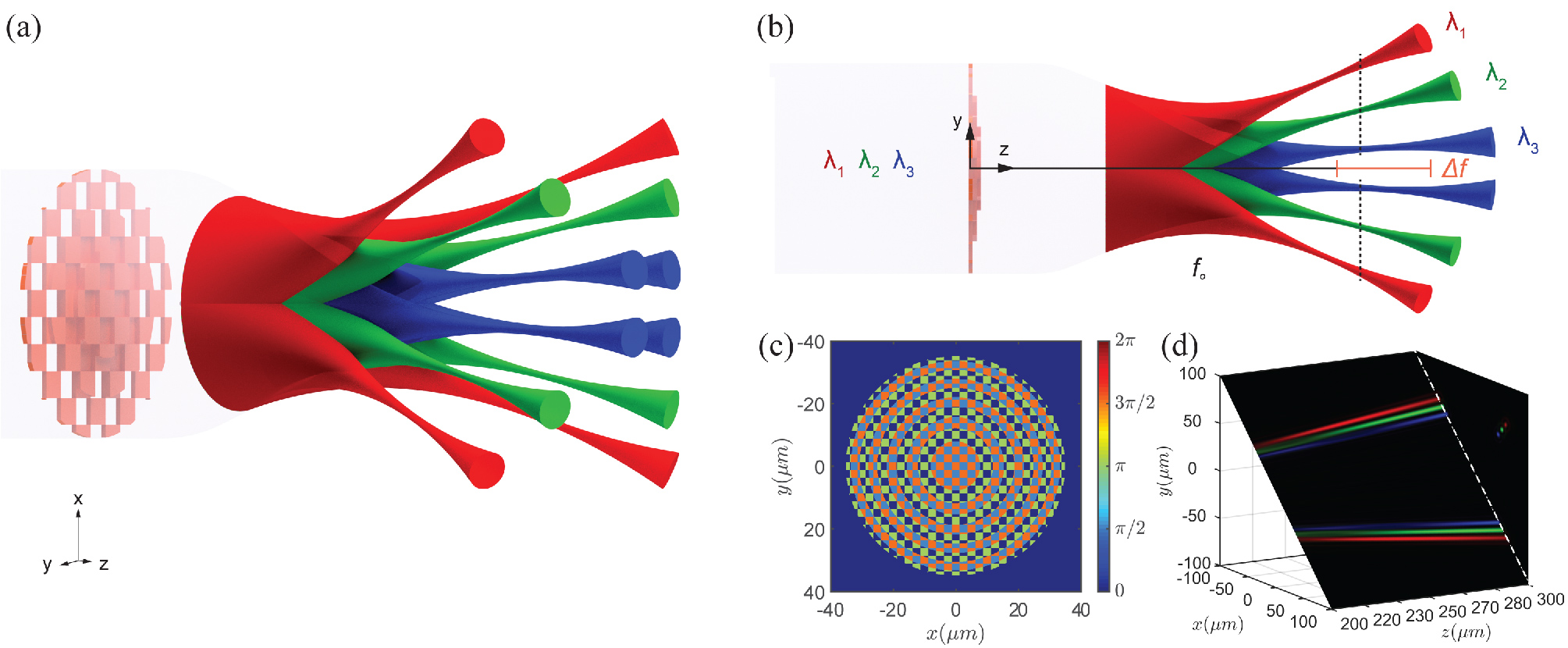}
\caption{Panel (a) shows the three-dimensional (3D) schematics of the phase-modulated axilens device illuminated with radiation of three different wavelengths, $\lambda_1$, $\lambda_2$, and $\lambda_3$. This device simultaneously splits and focuses the incoming wavelengths at distinct locations on the $xy$ focusing plane. Panel (b) shows the behavior of the designed device in a side-view schematics that emphasises how its extended focal depth $\Delta f$ enables focusing of different wavelengths on a single achromatic plane---indicated by the dashed lines. In panel (c) we show the 4-level discretized phase profile of a 2D periodically-modulated axilens with a 35 $\mu$m total device radius. Panel (d) shows the intensity distribution on a $45^\circ$ plane of the radiation focused by the phase profile shown in panel (c) at the three wavelengths: 750nm (blue), 850nm (green), and 950nm (red).}
\label{Fig1}
\end{figure*}

Before introducing our novel designs, we review the basic operation principle of an axilens, which has a phase profile given by \cite{davidson1991holographic}:
\begin{equation}\label{phase profile of axilens}
\phi(r)=-\frac{2\pi}{\lambda}\left.\left[\sqrt{\left(f_0+\frac{r{\Delta}f}{R}\right)^2+r^2}-\left(f_0+\frac{r{\Delta}f}{R}\right)\right]\right\vert_{2\pi}
\end{equation}
where $r=\sqrt{x^2+y^2}$, $f_0$ is the focal length, and ${\Delta}f$ is the focal depth. The $2\pi$ subscript indicates that the phase is reduced by modulo $2\pi$. Differently from the focusing behavior of a traditional Fresnel lens, axilenses are characterized by a larger focal depth that can be controlled by changing the parameter ${\Delta}f$. Although different wavelengths are focused at different positions along the $z$-axis, the larger focal depth of axilenses compared to traditional lenses results in a substantial overlap of the focused intensities of different wavelengths on the same plane, thus establishing an achromatic focal plane \cite{chen2020phasemodulated}. We will build on this distinctive property of axilenses later in the manuscript to demonstrate single-lens spectrometers based on phase-modulated axilenses. 
                        
In order to enable spectroscopic functionalities within an ultra-compact design we propose to combine the phase profile of axilenses with additional transverse phase modulation functions such as chirped or periodic ones \cite{axilensOL}. In this work, we demonstrate a single-lens spectrometer using axilenses with periodic phase modulations. The resulting phase-modulated axilenses extend the reach of  multifunctional DOEs, which are used for beam generation and design \cite{vijayakumar2017design,vasara1989realization,vijayakumar2015conical}, and provide novel opportunities for multispectral imaging and spectroscopy. 

Fig \ref{Fig1} (a) illustrates the concept of a  phase-modulated axilens illuminated with radiation of three different wavelengths, $\lambda_1$, $\lambda_2$, and $\lambda_3$. This device angularly separates all the incident wavelengths and simultaneously focus them at different locations across the same $xy$-plane. We schematize in Fig. \ref{Fig1} (b) the side-view of panel (a) demonstrating how the enhanced depth of focusing $\Delta f$ of an axilens enables the simultaneous focusing of well-separated wavelengths on a single achromatic image plane, which we highlight by the black dashed line. Note that a desired focal depth ${\Delta}f$ can be obtained using an axilens regardless of the wavelength of the incident radiation \cite{vijayakumar2017design}. 

In Fig. \ref{Fig1} (c), we show the phase distribution of a modulated axilens that combines a 2D transverse periodic phase modulation with the focusing one. The phase profile parameters of the axilens were chosen as follows: $f_0=100{\mu}m$, ${\Delta}f=200{\mu}m$, ${\lambda}=850nm$. These parameters can be flexibly changed to address other desired spectral ranges and/or dielectric substrate thickness and the refractive indices of different materials. In Fig.\ref{Fig1}(d) we show the intensity distribution of the phase-modulated axilens sampled on a $45^\circ$ tilted plane and computed using the RS formulation according to 
the following expressions \cite{goodman2005introduction}:
\begin{eqnarray}\label{RS equation}
    U_{2}\left(x,y\right)=U_{1}\left(x^\prime,y^\prime\right) * h(x^\prime,y^\prime)\\
    h(x^\prime,y^\prime)=\frac{1}{2\pi}\frac{z}{r}  \left(\frac{1}{r}-jk\right)\frac{e^{\left(jkr\right)}}{r}
\end{eqnarray}
where $*$ denotes 2D space convolution, $U_1$, $U_2$ are the transverse field distributions in the object and image planes with coordinates $(x,y)$ and $(x^\prime,y^\prime)$, respectively. Moreover, $k$ is the incident wave number and $r=\sqrt{x^2+y^2+z^2}$, where $z$ is the distance between object and image plane.

\begin{figure*}[t!]
\centering
\includegraphics[width=\linewidth]{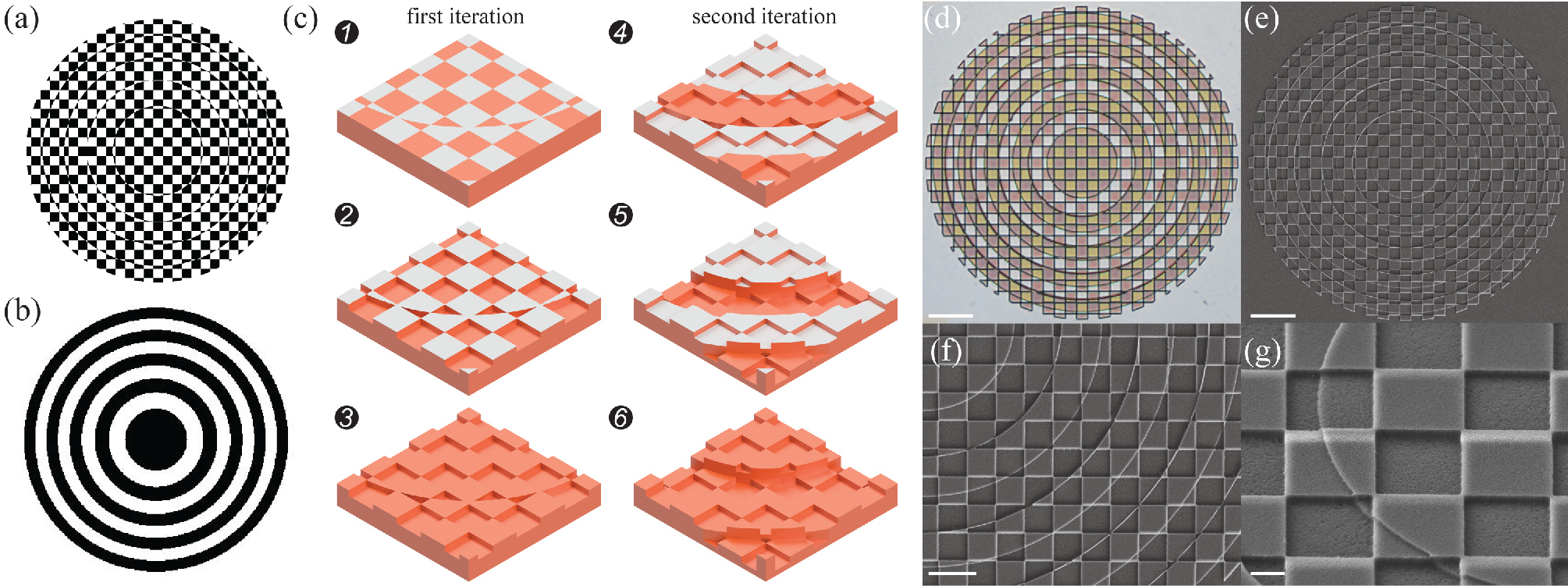}
\caption{(a-b) First and second lithographic processing iteration hardmask designs used for the phase-modulated axilens fabrication, respectively. (c) Process flow for four-level MDL fabrication with top-down lithographic methods. Steps 1-3 and 4-6 are associated with the first and second processing iterations, respectively. For both iterations, the first step is the lithographic patterning, metal deposition, and liftoff to form a hardmask; the second step is the anisotropic etching of exposed material; the third step is an etch that removes the residual hardmask. (d) Optical reflection bright field top-view image of a phase-modulated axilens fabricated by this method with a 10 $\mu$m scalebar. Color change at different material thicknesses is created by thin film interference effects. (e) SEM top-view image of the same device also with 10 $\mu$m scalebar. (f) Higher magnification SEM top-view image with a 4 $\mu$m scalebar. (g) SEM image angled at $45^\circ$ with a 2 $\mu$m scalebar.}
\label{Fig2}
\end{figure*}

Specifically, the incident plane waves interact with the phase profile shown in Fig. \ref{Fig1}(c) for the three wavelengths 750nm, 850nm, and 950nm and are both focused and diffracted into  well-separated beams that are clearly visible in the $45^\circ$ tilted plane while maintaining their focused character at the $z=300{\mu}m$ achromatic plane.

We experimentally demonstrate the designed 2D periodically-modulated ultra-compact expenses using reactively sputtered hydrogenated amorphous Si (a-Si:H). This material has excellent transparency and a high refractive index ($n\approx3.0$) at visible and NIR wavelengths. Likewise, physical deposition by sputtering is a low-cost and scalable process. Markedly, well-hydrogenated and alloyed low-loss a-Si:H is transparent until a sharp increase in the optical absorption edge appears around and 2.2 eV, depending on the specific deposition conditions \cite{moustakas1979sputtered}. The a-Si:H thin-films utilized in this study were grown reactively on fused silica, SiO$_2$, substrates using a Denton Discovery D8 magnetron sputtering system. Before deposition, substrates were solvent washed, sonicated, and plasma ashed in an O$_2$ atmosphere (M4L, PVA TePla America). The a-Si:H films were deposited using the following conditions: the sputtering base pressure was kept below $3\times10^{-7}$ Torr, the deposition pressure was kept at $10$ mTorr with a Ar:H$_2$ gas flow ratio of 4:1 Sccm, the substrate was heated to $300^\circ$C, and the RF deposition power on the $3''$ intrinsic Si target was set to 200W. The optical dispersion properties of the fabricated materials were characterized using spectroscopic ellipsometry (V-VASE, J.A. Woollam) on 50 nm thin-films. Our deposition process was optimized so that the final films used for fabrication have an average refractive index of $n=3.0$ and an optical transmission $\geq{85\%}$ across the entire operation range. 

Our fabrication method makes use of two lithographic processing steps to achieve a four-level discretized phase device. Figure 2(a-b) show the first and second etch mask designs, respectively. The first etch mask is created by combining the $3\pi/2$ and $\pi/2$ level phase areas, and the second etch mask is the combination of the $\pi$ and $3\pi/2$ level areas. The general fabrication process flow is shown in Fig. 2(c). Steps 1 through 3 and 4 through 6 are associated with the first and second processing iterations, respectively. First, steps 1 and 4 are the lithographic patterning of the first and second etch mask, respectively. The designed etch masks are transferred to metal hard masks with lithography and electron beam evaporation using a positive resist and a lift-off process. Next, steps 2 and 5 involve the definition of thicknesses $t = 1/4[\lambda_{0}/(n-1)]$ and $t = 1/2[\lambda_{0}/(n-1)]$ using deep dry etch, respectively. Finally, steps 3 and 6 consist of wet etch removals of the residual hard masks.

\begin{figure*}[t!]
\centering
\includegraphics[width=\linewidth]{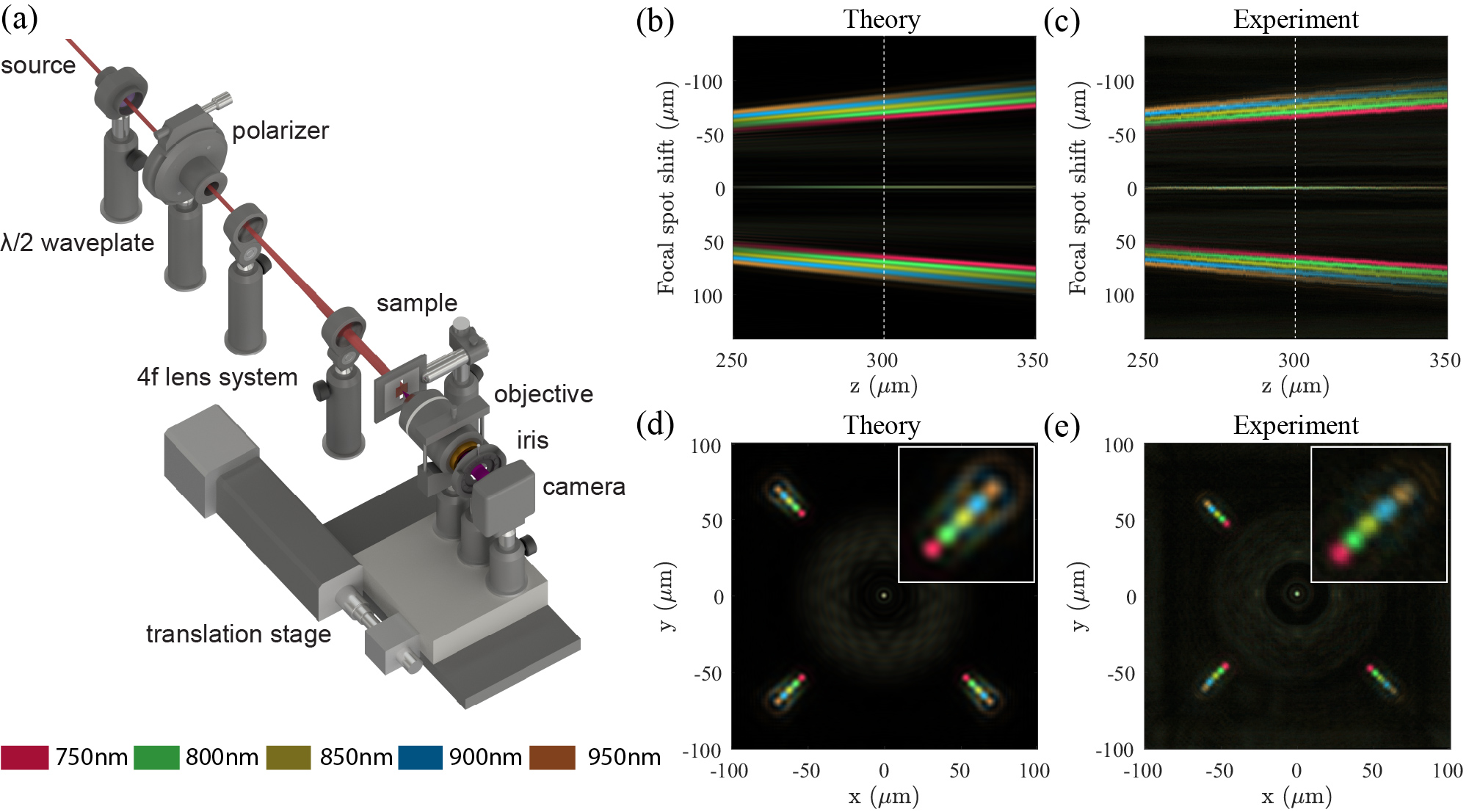}
\caption{(a) Experimental setup for measuring focusing character as a function of distance along the optical axis $z$. (b-c) Simulated and measured intensity profiles for five different wavelengths sampled at a $45^\circ$ plane (similar to \ref{Fig1}(d)), respectively.(d-e) Simulated and measured intensity profiles for the same five wavelengths on the designed achromatic focusing plane $z=300{\mu}m$, respectively.}
\label{Fig3}
\end{figure*}

The specific fabrication process demonstrated here utilizes electron beam lithography (EBL), a Cr hardmask, and a reactive ion etching (RIE) anisotropic dry etch for both processing iterations. In addition, we deposit Au alignment marks with Ti adhesion layers before device fabrication. For the lithography process, MicroChem PMMA A3 positive resist was spun at 3000 rpm and baked in a oven at 170$^\circ C$ for 20 min. After both resist baking steps, a thin conducting layer of Au was deposited using a Cressington 108 Manual Sputter Coater. The EBL alignment and exposure process was performed on a Zeiss Supra 40 scanning electron microscope (SEM) with NPGS writing software and a beam blanker. Exposure was at 30 keV with a 38 pA current and an area dosage of 250 $\mu$C/cm$^2$. Before development, the conductive layer was removed with a 3s Au wet etchant step. The resist was developed for 70s in 1:2 MIBK:IPA solution followed by a 20s immersion in IPA solution and DI-H$_2$O rinse. The Cr hardmasks (20 nm), Au alignment marks (30 nm), and Ti adhesion layer (10 nm) were deposited by electron beam evaporation using an Angstrom Engineering EVOVAC system. The anisotropic dry etching was performed with SF$_6$ gas at 150 W with a Plasma-Therm 790 RIE. The etch rate of the a-Si:H thin films under these conditions was found to be about 1.2 nm/s. The etch rate was calibrated using optical profilometry (Zygo New View 6300 Optical Surface Profiler) on a reference structure. The actual etch rate was measured to be reduced due to RIE lag. Lastly, the residual Cr hard mask was removed by a 6 min wet etch in Transene Inc. Cr Etchant 1020.

The fabricated device is in excellent agreement with the designed structure. Figure 2(d-e) show optical and SEM images of the top-view of the fabricated device. Each of the four phase layers in the optical image can be identified by the different colors due to color enhancement by optical interference. Higher magnification SEM images of the fabricated device are shown in Fig. 2(f-g). These images show all four levels with a total thickness of 318 nm and demonstrate the high-quality of the fabricated device.

The focusing behavior of the fabricated devices is experimentally characterized using the customized visible and near-infrared (NIR) imaging system with $z$-translation capabilities illustrated in Fig. 3(a). It utilizes a tunable Ti:sapphire (Mai Tai, Spectra-Physicsop) quasi-CW femtosecond laser source (82 MHz repetition rate, 100 fs pulse duration) over a broad visible and NIR spectral band (690 nm -- 1000 nm). We incorporate a $\lambda/2$ waveplate and a linear polarizer to attenuate both sources. In addition, a 4f lens system is used to expand the beam, which approximates a planewave source with negligible divergence. An imaging system consisting of a 0.9 numerical aperture (NA) objective (Olympus MPLFLN100xBDP), an iris (to remove stray reflections), and a CMOS camera (Thorlabs DCC1645C, with short-pass filter removed) is translated along the $z$-axis in unison by a motorized translation stage (Thorlabs LNR50S). This allows us to accurately characterize the intensity distribution in the achromatic focal plane as a function of the distance along the $z$-axis.

We show in Fig. \ref{Fig3} (b-c) the RS calculated and the measured intensity distributions, respectively, sampled on a $45^\circ$ tilted plane. The incident radiation at the different wavelengths has the same intensity. We used false colors to identify the different wavelengths as specified in the legend of Fig. \ref{Fig3}. We can observe that different incident wavelengths are all focused on the same achromatic plane, indicated by the white dashed line, 300 $\mu$m behind the device plane. In Figure \ref{Fig3} (d-e) we compare the calculated and the measured front-view of the intensity distributions corresponding to the different wavelengths on the achromatic plane (z=$300{\mu}m$), demonstrating clear spectral separation in very good agreement with the RS calculations.  
\begin{figure}[t]
\centering
\includegraphics[width=\linewidth]{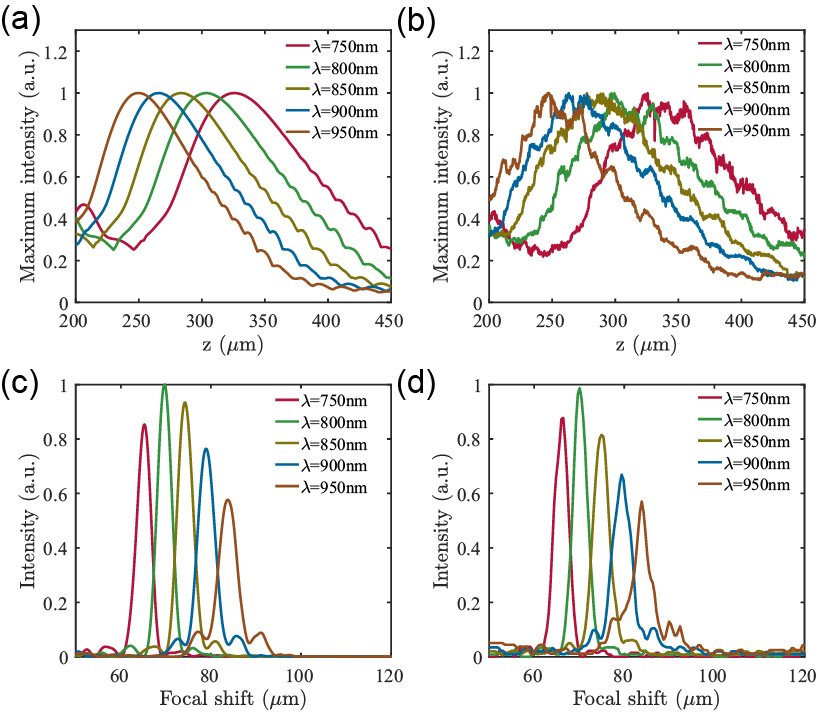}
\caption{Panels (a-b) show the normalized maximum intensity profiles at different distances along the optical z-axis according to numerical simulations and experimental data, respectively. Panels (c-d) show the normalized intensity profiles for different wavelengths on the achromatic plane $z=300{\mu}m$ along a line-cut from the center obtained from simulations and experimental measurements, respectively.}
\label{Fig4}
\end{figure}
To further characterize the achromatic focusing behavior of our device, in Fig.\ref{Fig4} (a-b) we compare the simulated and the measured normalized maximum intensity distributions at different wavelengths along the z-axis, respectively. The different intensity profiles display an almost constant $200{\mu}m$ focal depth (FWHM in panels (a-b)) and thus maintain a relative large intensity at the achromatic plane ($z=300{\mu}m$). In Figure \ref{Fig4} (c-d) we show the computed and experimentally measured transverse spatial intensity distributions on the achromatic plane ($z=300{\mu}m$) for the different incident wavelengths sampled along a line-cut on the $45^{\circ}$ tilted plane with respect to its origin (focal shift). The results demonstrate that our device can focus different wavelengths on the same plane and deflect them at different transverse positions. 
\begin{figure}[t!]
\centering
\includegraphics[width=\linewidth]{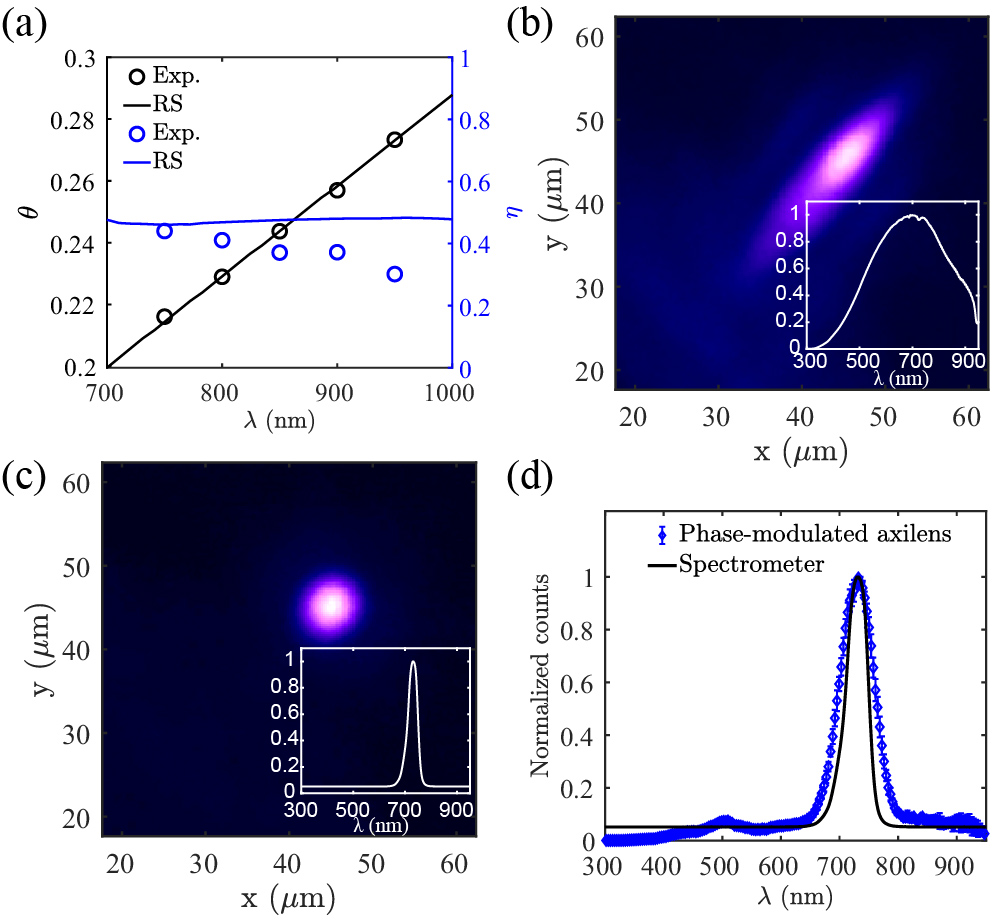}
\caption{Panel (a) shows the $\theta$ and $\eta$ as functions of incident wavelength from RS simulations and experiment measurements. Panel (b) shows the measured intensity distribution at $z=300{\mu}m$ when the device is illuminated with a halogen lamp used for intensity calibration. The inset shows the spectrum of the halogen lamp measured by a commercial spectrometer. Panel (c) shows the measured intensity distribution at $z=300{\mu}m$ when the device is illuminated by an LED source. Panel (d) shows the spectrum of the LED measured using our modulated axilens device and the commercial spectrometer for comparison.}
\label{Fig5}
\end{figure}

The angular dispersion and the focusing behavior of the fabricated devices is quantified in Fig. \ref{Fig5} (a) where we display the measured diffraction angle $\theta$, obtained from the spatial shift on the focal plane (at $z=300{\mu}m$) with respect to the center of the device, and of the focusing efficiency $\eta$. 
Our findings demonstrate linear control of the diffraction angle when increasing the incident radiation wavelength. The focusing efficiency $\eta$ is defined by the ratio with respect to the total incident power of the power localized within a region with a radius three times the full-width-at-half-maximum (FWHM) of the focal spot. We show that our device can focus the four spots with measured efficiency $\sim35\%$ over a broad range spanning from the visible to the near-infrared. From the slope of the linear dispersion data shown in Fig. \ref{Fig5} (a) we obtain the spectral resolution of the device  $\frac{d\theta}{d\lambda}=3.3*10^{-4}rad/m$. 
Finally, in Fig. \ref{Fig5} (b) and (c) we demonstrate that spatially modulated axilenses can be utilized as ultra-compact, single-lens spectrometers. In particular, we calibrated the dispersive response of the device using the broad-band spectrum $S_{halogen}(\lambda)$ of an halogen lamp, shown in the inset of panel (b), and then measured the emission spectrum of an LED source, shown in the inset of panel (c). Radiometric calibration of the modulated axilens was achieved by referencing the measured LED spectrum to the known spectrum of the halogen lamp, according to the following procedure:
first, we obtain the linear function $\ell(\lambda)$ that relates the focal shift $\ell$ along on the $45^{\circ}$ tilted achromatic plane to the incident wavelength from Fig.\ref{Fig5} (a). This is accomplished easily by considering $\ell=z\tan(\theta)$, where $z=300{\mu}m$ is the position of the achromatic plane, and $\theta$ is the diffraction angle shown in Fig.\ref{Fig5} (a). Second, we measure the intensity distributions $I_{halogen}(\ell)$ and $I_{LED}(\ell)$ along the $45^{\circ}$ plane for the halogen and the LED source, displayed in Fig.\ref{Fig5} (b-c), respectively. Third, we use the function $\ell(\lambda)$ to express the intensity distributions $I_{halogen}(\lambda)$ and $I_{LED}(\lambda)$ in the wavelength domain. Finally, we can obtain the correct LED source spectrum using the phase-modulated axilens using the following relation:
\begin{equation}
S_{LED}(\lambda)=I_{LED}(\lambda)\frac{S_{halogen}(\lambda)}{I_{halogen}(\lambda)}
\end{equation}

The signal collected from four different spots on the device was averaged and the results are shown in panel (d) where we compare with the emission  spectrum of the same LED independently measured with a commercial spectrometer system (Ocean Optics QE65000). The agreement between the two sets of data is quite remarkable. 
The minimum resolvable wavelength interval that we achieved with the current design of phase-modulated axilenses can be estimated using the relation \cite{demtroder2014laser}:
\begin{equation}    
    \Delta\lambda=\frac{\lambda_0}{D\frac{d\theta}{d\lambda_0}}
\end{equation}\label{minimum_interval}
where $\lambda_0$ is the wavelength in air and $D$ is the device diameter. This yields, for our choice of parameters, $\Delta\lambda=40nm$ at $\lambda_0=850nm$. Notice that $\Delta\lambda$ is inversely related to the device aperture size $D$. Therefore, miniaturized modulated axilens-based spectrometers necessarily have limited $\Delta\lambda$ compared to large-scale commercial spectrometers with a typical aperture size of $10~cm$. However, the $\Delta\lambda$ of phase-modulated axilenses can be further reduced by increasing the device diameter $D$ and/or further by reducing the periodicity $p$, depending on different application scenarios. For example, a modulated axilens-based microspectrometer with $D=200{\mu}m$ and a reduced grating period $p=2{\mu}m$, which can be fabricated using scalable photolithography, will display a much larger resolving power corresponding to a minimum resolvable wavelength interval $\Delta\lambda={5}nm$ at $\lambda=850{\mu}m$, enabling realistic spectroscopic applications on a miniaturized chip \cite{ReddingLiewSarmaCao:2013}.
Finally, we remark that current work using metasurfaces has demonstrated compact spectrometers in the visible range \cite{zhu2017ultra,zhu2019compact} that achieve 1 $nm$ spectral resolution over a 200 $nm$ bandwidth. However, such devices requires polarization control and feature reduced focusing efficiency of approximately $\sim12\%$. Besides, they operate with an off-axis focal plane that requires precision mounting that makes, for instance, the integration with FPAs quite challenging. In contrast, our proposed device is polarization-insensitivity, smaller in size and has a focusing efficiency $\sim 35\%$ over a bandwidth of $200nm$. Moreover, since it does not require off-axis focusing planes it can be readily integrated atop FPAs.

In conclusion, we have designed and demonstrated a 4-level phase-modulated axilens novel device platform that can be directly integrated atop FPAs to achieve imaging and spectroscopic  functionalities. Building on the concept of a phase-modulated axilens, we demonstrate and characterize the achromatic focusing and spectroscopic behavior of these devices over a broad spectral range. Specifically, we design and demonstrate ultra-compact (70$\mu$m diameter, $\approx$ 300nm thickness) proof-of-concept single-lens microspectrometers with 200 $nm$ bandwidth and $\sim 35\%$ focusing efficiency. We remark that the spectral resolution and $\Delta \lambda$ of these devices is not limited to the value provided above and can be further improved to $\Delta \lambda=5nm$ using current lithographic techniques. The structures demonstrated in this paper add fundamental imaging and spectroscopic capabilities to traditional DOEs and represent alternative solutions, with respect to metasurface-based designs, for critical applications such as spectral sorting, multi-band IR imaging, photodetection, and spectroscopy.

\section*{funding}{This research was sponsored by the Army Research Laboratory and was accomplished under Cooperative Agreement Number W911NF-12-2-0023. The views and conclusions contained in this document are those of the authors and should not be interpreted as representing the official policies, either expressed or implied, of the Army Research Laboratory or the U.S. Government. The U.S. Government is authorized to reproduce and distribute reprints for Government purposes notwithstanding any copyright notation herein.}

\section*{Acknowledgements}
L.D.N. would like to thank Prof. Enrico Bellotti for insightful discussions on the presented technology.


%
\end{document}